\documentclass[twocolumn, a4paper, superscriptaddress,showpacs,amsmath,amssymb,aps,prc,floatfix,superscriptaddress]{revtex4}
\usepackage{graphicx}
\usepackage{dcolumn}
\usepackage{bm}
\usepackage{multirow}
\usepackage{amsmath}
\usepackage{longtable}
\usepackage{epstopdf}

\begin{document}

\title{Simultaneous investigation of the  $\mathbf{T=1~ (J^{\pi}=0^+)}$ and  $\mathbf{T=0 ~ (J^{\pi}=9^+)}$ $\beta$ decays in $^{70}$Br}
\author{A.I.~Morales}
\email{Ana.Morales@ific.uv.es}
\affiliation{IFIC, CSIC-Universitat de Val\`encia, E-46071 Val\`encia, Spain}
\affiliation{Dipartimento di Fisica dell'Universit\`a degli Studi di Milano, I-20133 Milano, Italy}
\affiliation{Istituto Nazionale di Fisica Nucleare, Sezione di Milano, I-20133 Milano, Italy}

\author{A.~Algora}
\email{algora@ific.uv.es}
\affiliation{IFIC, CSIC-Universitat de Val\`encia, E-46071 Val\`encia, Spain}
\affiliation{Institute of Nuclear Research of the Hungarian Academy of Sciences, H-4001 Debrecen, Hungary}

\author{B.~Rubio}
\affiliation{IFIC, CSIC-Universitat de Val\`encia, E-46071 Val\`encia, Spain}

\author{K.~Kaneko}
\affiliation{Department of Physics, Kyushu Sangyo University, Fukuoka 813-8503, Japan}

\author{S.~Nishimura}
\affiliation{RIKEN Nishina Center, 2-1 Hirosawa, Wako, Saitama 351-0198, Japan}

\author{P.~Aguilera}
\affiliation{Comisi\'on Chilena de Energ\'ia Nuclear, Casilla 188-D, Amun\'ategui 95, Santiago Centro, Santiago, Chile}

\author{S.E.A.~Orrigo}
\affiliation{IFIC, CSIC-Universitat de Val\`encia, E-46071 Val\`encia, Spain}

\author{F.~Molina}
\affiliation{Comisi\'on Chilena de Energ\'ia Nuclear, Casilla 188-D, Amun\'ategui 95, Santiago Centro, Santiago, Chile}

\author{G.~de~Angelis}
\affiliation{Istituto Nazionale di Fisica Nucleare, Laboratori Nazionali di Legnaro, Legnaro, Italy}

\author{F.~Recchia}
\affiliation{Dipartimento di Fisica dell'Universit\`a degli Studi di Padova, Padova, Italy}
\affiliation{Istituto Nazionale di Fisica Nucleare, Sezione di Padova, Padova, Italy}

\author{G.~Kiss}
\affiliation{RIKEN Nishina Center, 2-1 Hirosawa, Wako, Saitama 351-0198, Japan}

\author{V. H. Phong}
\affiliation{RIKEN Nishina Center, 2-1 Hirosawa, Wako, Saitama 351-0198, Japan}
\affiliation{Faculty of Physics, VNU Hanoi University of Science, 334 Nguyen Trai, Thanh Xuan, Hanoi, Vietnam}
 
\author{J.~Wu}
\affiliation{RIKEN Nishina Center, 2-1 Hirosawa, Wako, Saitama 351-0198, Japan}

\author{D.~Nishimura}
\affiliation{Department of Physics, Tokyo City University, Setagaya-ku, Tokyo 158-8557, Japan}

\author{H.~Oikawa}
\affiliation{Department of Physics, Tokyo University of Science, Noda, Chiba 278-8510, Japan}

\author{T.~Goigoux}
\affiliation{Centre d'Etudes Nucl\'eaires de Bordeaux-Gradignan - UMR 5797 Universit\'e de Bordeaux
CNRS/IN2P3, 19 Chemin du Solarium, CS10120, 33175 Gradignan Cedex, France}

\author{J.~Giovinazzo}
\affiliation{Centre d'Etudes Nucl\'eaires de Bordeaux-Gradignan - UMR 5797 Universit\'e de Bordeaux
CNRS/IN2P3, 19 Chemin du Solarium, CS10120, 33175 Gradignan Cedex, France}

\author{P.~Ascher}
\affiliation{Centre d'Etudes Nucl\'eaires de Bordeaux-Gradignan - UMR 5797 Universit\'e de Bordeaux
CNRS/IN2P3, 19 Chemin du Solarium, CS10120, 33175 Gradignan Cedex, France}

\author{J.~Agramunt}
\affiliation{IFIC, CSIC-Universitat de Val\`encia, E-46071 Val\`encia, Spain}

\author{D.S.~Ahn}
\affiliation{RIKEN Nishina Center, 2-1 Hirosawa, Wako, Saitama 351-0198, Japan}

\author{H.~Baba}
\affiliation{RIKEN Nishina Center, 2-1 Hirosawa, Wako, Saitama 351-0198, Japan}

\author{B.~Blank}
\affiliation{Centre d'Etudes Nucl\'eaires de Bordeaux-Gradignan - UMR 5797 Universit\'e de Bordeaux
CNRS/IN2P3, 19 Chemin du Solarium, CS10120, 33175 Gradignan Cedex, France}

\author{C.~Borcea}
\affiliation{National Institute for Physics and Nuclear Engineering IFIN-HH, P.O. Box MG-6, Bucharest-Magurele, Romania}

\author{A.~Boso}
\affiliation{Dipartimento di Fisica dell'Universit\`a degli Studi di Padova, Padova, Italy}
\affiliation{Istituto Nazionale di Fisica Nucleare, Sezione di Padova, Padova, Italy}

\author{P.~Davies}
\affiliation{Department of Physics, University of York, Heslington, York YO10 5DD, United Kingdom}

\author{F.~Diel}
\affiliation{Institute of Nuclear Physics, University of Cologne, D-50937 Cologne, Germany}

\author{Zs.~Dombr\'adi}
\affiliation{Institute of Nuclear Research of the Hungarian Academy of Sciences, H-4001 Debrecen, Hungary}

\author{P.~Doornenbal}
\affiliation{RIKEN Nishina Center, 2-1 Hirosawa, Wako, Saitama 351-0198, Japan}

\author{J.~Eberth}
\affiliation{Institute of Nuclear Physics, University of Cologne, D-50937 Cologne, Germany}

\author{G.~de~France}
\affiliation{Grand Acc\'el\'erateur National d'Ions Lourds, B.P. 55027, F-14076 Caen Cedex 05, France}

\author{Y.~Fujita}
\affiliation{Department of Physics, Osaka University, Toyonaka, Osaka 560-0043, Japan}

\author{N.~Fukuda}
\affiliation{RIKEN Nishina Center, 2-1 Hirosawa, Wako, Saitama 351-0198, Japan}

\author{E.~Ganioglu}
\affiliation{Department of Physics, Istanbul University, Istanbul 34134, Turkey}

\author{W.~Gelletly}
\affiliation{IFIC, CSIC-Universitat de Val\`encia, E-46071 Val\`encia, Spain}
\affiliation{Department of Physics, University of Surrey, Guildford GU2 7XH, United Kingdom}

\author{M.~Gerbaux}
\affiliation{Centre d'Etudes Nucl\'eaires de Bordeaux-Gradignan - UMR 5797 Universit\'e de Bordeaux
CNRS/IN2P3, 19 Chemin du Solarium, CS10120, 33175 Gradignan Cedex, France}

\author{S.~Gr\'evy}
\affiliation{Centre d'Etudes Nucl\'eaires de Bordeaux-Gradignan - UMR 5797 Universit\'e de Bordeaux
CNRS/IN2P3, 19 Chemin du Solarium, CS10120, 33175 Gradignan Cedex, France}

\author{V.~Guadilla}
\affiliation{IFIC, CSIC-Universitat de Val\`encia, E-46071 Val\`encia, Spain}

\author{N.~Inabe}
\affiliation{RIKEN Nishina Center, 2-1 Hirosawa, Wako, Saitama 351-0198, Japan}

\author{T.~Isobe}
\affiliation{RIKEN Nishina Center, 2-1 Hirosawa, Wako, Saitama 351-0198, Japan}

\author{W.~Korten}
\affiliation{CEA Saclay, IRFU, SPhN, 91191 Gif-sur-Yvette, France}

\author{T.~Kubo}
\affiliation{RIKEN Nishina Center, 2-1 Hirosawa, Wako, Saitama 351-0198, Japan}

\author{S.~Kubono}
\affiliation{RIKEN Nishina Center, 2-1 Hirosawa, Wako, Saitama 351-0198, Japan}

\author{T.~Kurtuki\'an Nieto}
\affiliation{Centre d'Etudes Nucl\'eaires de Bordeaux-Gradignan - UMR 5797 Universit\'e de Bordeaux
CNRS/IN2P3, 19 Chemin du Solarium, CS10120, 33175 Gradignan Cedex, France}

\author{S.~Lenzi}
\affiliation{Dipartimento di Fisica dell'Universit\`a degli Studi di Padova, Padova, Italy}
\affiliation{Istituto Nazionale di Fisica Nucleare, Sezione di Padova, Padova, Italy}

\author{D.~Lubos}
\affiliation{Physik Department E12, Technische Universit\"at M\"unchen, D-85748 Garching, Germany}

\author{C.~Magron}
\affiliation{Centre d'Etudes Nucl\'eaires de Bordeaux-Gradignan - UMR 5797 Universit\'e de Bordeaux
CNRS/IN2P3, 19 Chemin du Solarium, CS10120, 33175 Gradignan Cedex, France}

\author{A.~Montaner-Piz\'a}
\affiliation{IFIC, CSIC-Universitat de Val\`encia, E-46071 Val\`encia, Spain}

\author{D.R.~Napoli}
\affiliation{Dipartimento di Fisica dell'Universit\`a degli Studi di Padova, Padova, Italy}
\affiliation{Istituto Nazionale di Fisica Nucleare, Sezione di Padova, Padova, Italy}

\author{H.~Sakurai}
\affiliation{RIKEN Nishina Center, 2-1 Hirosawa, Wako, Saitama 351-0198, Japan}
\affiliation{Department of Physics, Tohoku University, Miyagi 980-8578, Japan}

\author{Y.~Shimizu}
\affiliation{RIKEN Nishina Center, 2-1 Hirosawa, Wako, Saitama 351-0198, Japan}

\author{C.~Sidong}
\affiliation{RIKEN Nishina Center, 2-1 Hirosawa, Wako, Saitama 351-0198, Japan}

\author{P.-A.~S\"oderstr\"om}%
\affiliation{RIKEN Nishina Center, 2-1 Hirosawa, Wako, Saitama 351-0198, Japan}

\author{T.~Sumikama}%
\affiliation{Department of Physics, Tohoku University, Miyagi 980-8578, Japan}

\author{H.~Suzuki}
\affiliation{RIKEN Nishina Center, 2-1 Hirosawa, Wako, Saitama 351-0198, Japan}

\author{H.~Takeda}
\affiliation{RIKEN Nishina Center, 2-1 Hirosawa, Wako, Saitama 351-0198, Japan}  

\author{Y.~Takei}
\affiliation{Department of Physics, Tokyo University of Science, Noda, Chiba 278-8510, Japan}

\author{M.~Tanaka}
\affiliation{Department of Physics, Osaka University, Toyonaka, Osaka 560-0043, Japan}

\author{S.~Yagi}
\affiliation{Department of Physics, Tokyo University of Science, Noda, Chiba 278-8510, Japan}

\date{\today}

\begin{abstract}

The $\beta$ decay of the odd-odd nucleus $^{70}$Br has been investigated with the BigRIPS and EURICA setups at the Radioactive Ion Beam Factory (RIBF) of the RIKEN Nishina Center. The $T=0$ ($J^{\pi}=9^+$) and $T=1$ ($J^{\pi}=0^+$) isomers have both been produced in in-flight fragmentation of $^{78}$Kr with ratios of 41.6(8)\%  and 58.4(8)\%, respectively. A half-life of $t_{1/2}=2157^{+53}_{-49}$ ms has been measured for the $J^{\pi}=9^+$ isomer from $\gamma$-ray time decay analysis.  Based on this result, we provide a new value of the half-life for the $J^{\pi}=0^+$ ground state of $^{70}$Br, $t_{1/2}=78.42\pm0.51$ ms, which is slightly more precise, and in excellent agreement, with the best measurement reported hitherto in the literature. For this decay, we provide the first estimate of the total branching fraction decaying through the $2^+_1$ state in the daughter nucleus $^{70}$Se, $R(2^+_1)=1.3\pm1.1\%$. We also report four new low-intensity $\gamma$-ray transitions at 661, 1103, 1561, and 1749 keV following the $\beta$ decay of the $J^{\pi}=9^+$ isomer. Based on their coincidence relationships, we tentatively propose two new excited states at  3945 and 4752 keV in $^{70}$Se with most probable spins and parities of $J^{\pi}=(6^+)$ and $(8^+)$, respectively. The observed structure is interpreted with the help of shell-model calculations, which predict a complex interplay between oblate and prolate configurations at low excitation energies.   

\end{abstract}


\maketitle

\section{Introduction}
\label{sec:intro}

One of the core concepts of the electroweak standard model is the unitarity of the Cabibbo-Kobayashi-Maskawa (CKM) matrix which is used to describe the quark weak-interaction eigenstates in terms of the quark mass eigenstates \cite{Har13}. Hitherto, superallowed $\beta$ transitions between $J^{\pi}=0^+$, $T=1$ analog states have provided the most precise value of the largest matrix element, the up-down term $V_{ud}$. $V_{ud}$ can be extracted from the ratio between $G_V$, the vector coupling constant for a semileptonic decay, and $G_F$, the weak-interaction constant for a pure leptonic decay. The Conserved Vector Current (CVC) hypothesis postulates that $G_V$ is a universal constant independent of the nuclear medium. This means that the strength or $ft$ value of the superallowed Fermi transitions, which are only mediated by the vector current, is the same for nuclei with identical isospin. As a result, a mean $ft$ value can be used to determine $V_{ud}$ \cite{Har15}. 

Experimental $ft$ values are obtained from the total transition energy $Q_{EC}$ (required for the calculation of $f$), the half-life of the parent state, and the branching ratio of the superallowed Fermi transition (both required for the calculation of $t$). In practice, small corrections accounting for radiative and isospin symmetry-breaking effects are incorporated  as discussed in Ref. \cite{Har15}, resulting in a ``corrected'' $\mathcal{F}t$ value given by

\begin{eqnarray}
\mathcal{F}t \equiv ft(1+\delta^{'}_{R})(1+\delta_{NS}-\delta_{C})=\frac{K}{2G^{2}_{V}(1+\Delta^{V}_{R})}
 \label{eq:eq0}
\end{eqnarray}

where $K$ is a constant, $\delta_{C}$ is the isospin symmetry-breaking correction,  $\Delta^{V}_{R}$ is the transition-independent part of the radiative correction, and $\delta^{'}_{R}$ and 
$\delta_{NS}$ are the transition dependent parts of the radiative correction \cite{Har15}. 

Up to now, 14 superallowed $\beta$ decays have been measured with enough precision to test the CVC hypothesis, resulting in a world-average corrected $\mathcal{F}t=3072.27\pm0.72$ s \cite{Har15}.  The superallowed $\beta$ decay of $^{70}$Br has been excluded from this compilation due to conflicting experimental values for the $Q_{EC}$ decay energy \cite{Dav80,Sav09}. On top of this, the currently adopted half-life is only known to a precision of $\sim10$ parts in a thousand and the superallowed $0^+\rightarrow0^+$ branching ratio has not been measured yet \cite{Har15}. This decay, combined with the relatively recent results on $^{62}$Ga \cite{Fin08} and $^{74}$Rb \cite{Dun13}, is an excellent testing ground for theoretical modeling due to the increased role of the charge-dependent corrections in nuclei with A $\geq$ 62 \cite{Har02}. 
   
In addition to the CVC motivation, the nuclei in this mass region are of particular interest because their properties change rapidly with the addition or subtraction of one nucleon.  The resulting scenario of nuclear shape evolution is intricate and challenges state-of-the-art nuclear models, which have to cope with the development of spherical, oblate, and prolate shapes stabilized by the occupation of the deformation-driving orbital $g_{9/2}$ \cite{Has07,Mol09,Bai15}. Moreover different shape minima may compete simultaneously at low spins and excitation energies in a single nucleus, leading to the occurrence of shape coexistence. This phenomenon, widely studied in recent years (see Ref. \cite{Hey11} for a review), may help us to understand the microscopic mechanisms enhancing quantum many-body correlations in exotic nuclei \cite{Ots16,Kre16,Mor17}.

Experimentally, the picture of shape-related phenomena in the Se (Z=34) isotopic chain is more ambiguous than for the heavier Kr (Z=36) isotopes \cite{Gad05,Cle07,Poi04,Bri15,Gor05,Iwa14}. It has been shown that the ground states of the even-even $^{74-82}$Se are prolate-deformed \cite{Lec77,Lec78}. However, for the lighter $^{68,70,72}$Se isotopes, there is evidence for oblate-deformed ground states \cite{Fis00,Lju08,McC11}, but this has been called into question by other studies \cite{Hur07,Obe09}. Theoretically, a number of microscopic approaches support an oblate ground-state shape that evolves rapidly into a prolate collective rotation for spin $J\geqslant6$  \cite{Myl89,Pet03,Kan04,Lju08,Hin09,Kan15,Pet15}. This is consistent with the measured moments of inertia (see, for instance, Fig. 1 of Ref. \cite{McC11}). There are discrepancies, though, in the value of the spin at which the phase-shape transition occurs. This is due to the differences found between models in the predicted shape minima and configuration mixing, which lead to different shape-coexistence pictures. Hence, further experimental information on the competing shapes is important to completely understand the rapid shape changes in the $A\sim70$ mass region. 

It is worth also noting that most of these nuclei are on the pathway of the  \textit{rp}-process of stellar nucleosynthesis \cite{Wal81,Sch01,Sav09} and some of them, as $^{68}$Se and $^{72}$Kr, are important waiting-points defining the time scale of the process. Here, shape coexistence and mixing are crucial to determine the shape of the $\beta$-strength distribution and, hence, the stellar $\beta$-decay rates. 

In the present article we report on a $\beta$-decay study of the odd-odd self-conjugate nucleus $^{70}$Br. The experiment was performed at the RIKEN Nishina Center using the BigRIPS spectrometer and the EURICA $\beta$-decay station \cite{Nis12}. The results are of relevance in terms of deriving an additional $\mathcal{F}t$ value for the CVC test and extending our knowledge of the structure of $^{70}$Se. We present here an improved measurement of the half-life of the $T=1$ ($J^{\pi}=0^+$) ground state of $^{70}$Br and, for the first time, an estimate of the total branching fraction to the $2^+_1$ state in the daughter nucleus $^{70}$Se. 

In previous half-life measurements, $^{70}$Br nuclei were produced in fragmentation and fusion-evaporation reactions \cite{Alb78,Bur88,Lop02,Rog14}, which are optimal tools for the population of isomeric states. Though the $T=0 ~ (J^{\pi}=9^+)$ $\beta$-decaying isomer was already known by the time most of these works were published \cite{Vos81}, its contribution to the measured decay curve has hitherto been neglected. Therefore, this is the first work in which both the  $T=1~ (J^{\pi}=0^+)$ and  $T=0 ~ (J^{\pi}=9^+)$ $\beta$ decays of $^{70}$Br were investigated simultaneously. 

The paper is organized as follows. In Sec. \ref{sec:exp} the experimental apparatus is presented. Section \ref{sec:corr} describes the correlations defined to obtain the $\beta$-decay information. Next we present the results for the $T=0~(J^{\pi}=9^+)$ (Sec. \ref{sec:long}) and  $T=1~(J^{\pi}=0^+)$ (Sec. \ref{sec:short}) states of $^{70}$Br. A discussion of these results, including a comparison with shell-model calculations, is provided in Sec. \ref{sec:discussion}. Finally, we present a brief summary and conclusions in Sec. \ref{sec:conc}.

\section{The experimental setup}
\label{sec:exp}

$^{70}$Br nuclei were produced in the fragmentation of relativistic $^{78}$Kr projectiles impinging on a 5-mm thick Be target. The $^{78}$Kr primary beam was delivered by the RILAC2-RRC-fRC-IRC-SRC acceleration system with an energy of 345 MeV/nucleon and an average intensity of 38 pnA \cite{Nis16}. The exotic fragmentation residues were separated and selected in the BigRIPS separator using the $\Delta E-B\rho-ToF$ method, which is based on an event-by-event measurement of the energy loss ($\Delta E$), magnetic rigidity ($B\rho$), and time-of-flight ($ToF$) of the ions. Multi-sampling ionization chambers, parallel-plate avalanche counters, and plastic scintillators were located in the focal planes of the beam line for this purpose. Fragmentation products were then identified by their mass-to-charge ratio (A/Q) and atomic number (Z) using standard particle-identification procedures \cite{Fuk13,Bla16}. 

The nuclei were transported in-flight to a $\beta$-decay station located at the exit of the ZeroDegree spectrometer \cite{zerodegree}. The setup consisted of the Wide-range Active Silicon Strip Stopper Array for Beta and ion detection (WAS3ABi), used to measure the energy, time, and position of implantations and $\beta$ particles, and the EUroball-RIKEN Cluster Array (EURICA), used for the measurement of the energy and time of $\gamma$-ray transitions following implantation or $\beta$ decay \cite{Nis12,Sod13}. 

The nuclei were implanted in WAS3ABi with the help of a homogeneous aluminum degrader 1.2 mm thick. A plastic scintillator (henceforth called the veto detector) was located behind WAS3ABi to identify ions passing through the active stopper. In the present experiment, WAS3ABi consisted of a compact array of 3 double-sided silicon-strip detectors (DSSSD), each with $6\times4$ cm$^2$ area and 1 mm thickness. The DSSSD detectors were segmented in 60 vertical (X) and 40 horizontal (Y) strips, providing 2400 pixels with an active area of $1\times1$ mm$^2$ each.  Energy and time signals from each strip were read by standard analogue electronics in a high-gain branch suited for the energy range of decay particles. The energy range of the X strips was set to 4 MeV to optimize the energy resolution for $\beta$ particles, while that of the Y strips was extended to 10 MeV to detect high-energy protons. In-flight fragments reaching WAS3ABi with energies well above 1 GeV were registered as overflow signals. Since high-energy ions passing through WAS3ABi typically overflowed more than one strip, their implantation position was determined from the identification of the X and Y strips with the fastest time signals \cite{Nis13}.

The EURICA  $\gamma$-ray array was composed of 84 HPGe detectors grouped in 12 clusters of 7 crystals each. The clusters were placed surrounding WAS3ABi at an average distance of 22 cm, achieving a total absolute detection efficiency of 10\% at 662 keV. The $\gamma$ rays were registered up to 100 $\mu s$ after the detection of a triggering signal in WAS3ABi. This allowed for the detection of isomeric states ranging from several ns up to some ms in both parent and daughter nuclei. 

\begin{figure}
\vspace*{1cm}
\hspace*{-0.2cm}
\includegraphics[width=8 cm]{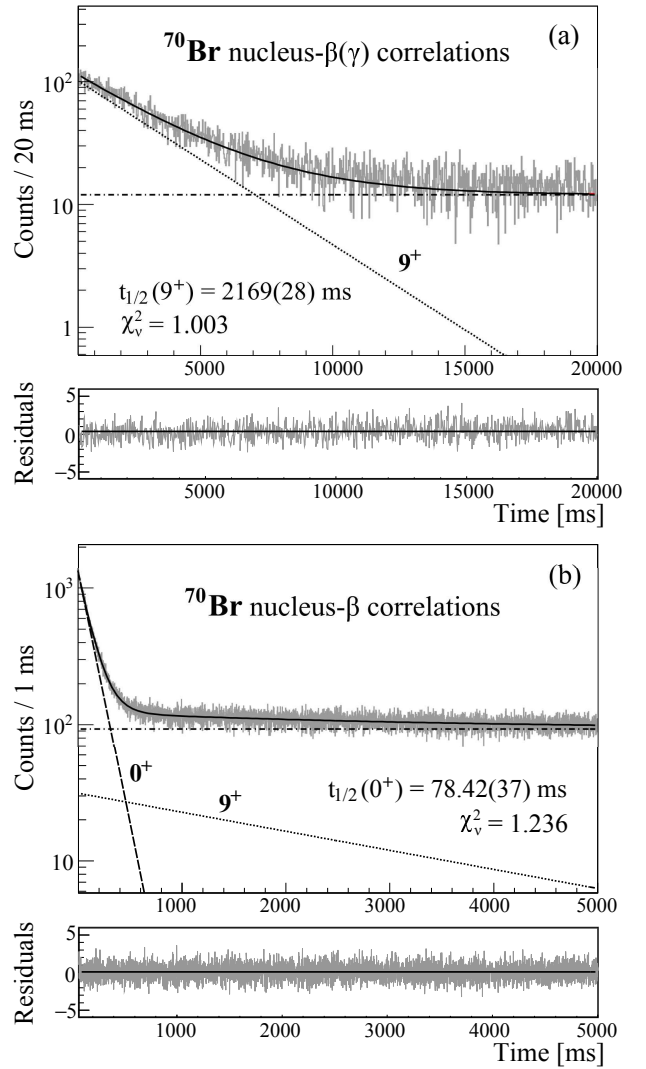}  
\caption{$\beta$-decay time spectra of $^{70}$Br including (a) ion-$\beta(\gamma)$ correlations. The time distributions of the 945-, 964-, 1034-, and 1094-keV $\gamma$ transitions, attributed to the decay of the $T=0~(J^{\pi}=9^+)$ isomer, are summed (b) ion-$\beta$ correlations with no $\gamma$-ray conditions. The activities of the ground state and $9^+$ isomer are indicated as dashed and dotted lines, respectively, while the background is shown as a dotted-dashed line. The fitting functions are indicated as continuous lines. The residual plots are presented at the bottom of each panel. The half-life and reduced $\chi^2$ obtained from each spectrum is shown together with the uncertainty from the fit.}
\label{fig1}
\end{figure}

\section{Correlation procedure}
\label{sec:corr}

Implanted nuclei and $\beta$ particles were identified and separated offline. Implantations were defined by (1) a high-energy signal in the last fast-plastic scintillators of the BigRIPS and the ZeroDegree spectrometers, (2) an overflow energy signal in WAS3ABi, and (3) no energy signal in the veto detector. On these criteria, about 1.3$\times10^6$ $^{70}$Br implantation events were registered in the central DSSSD of WAS3ABi. The implantation position was determined from the identification of the X and Y strips with the fastest time signal as described before. For each valid implantation, $\beta$ particles were accepted in correlation if (a) no high-energy signal was registered in the last fast-plastic scintillators of the BigRIPS and the ZeroDegree spectrometers, (b) no overflow energy signal was registered in the Y strips of WAS3ABi, (c) an energy signal above a variable threshold (see discussion in Sec. \ref{sec:short}A)  was detected in WAS3ABi in the same pixel as the implantation event, and (d) the time elapsed between the implantation and all subsequent $\beta$ particles was shorter than 20 s. The energy released by $\beta$-like events was obtained by adding the energies of neighboring strips in each DSSSD detector and their position was defined as the energy-weighted mean of the X and Y strips added together. The resulting activities were sorted as a function of time to extract $\beta$-decay half-lives. 

\begin{figure}
\vspace*{0.2cm}
\hspace*{-0.5cm}
\includegraphics[width=8.3 cm]{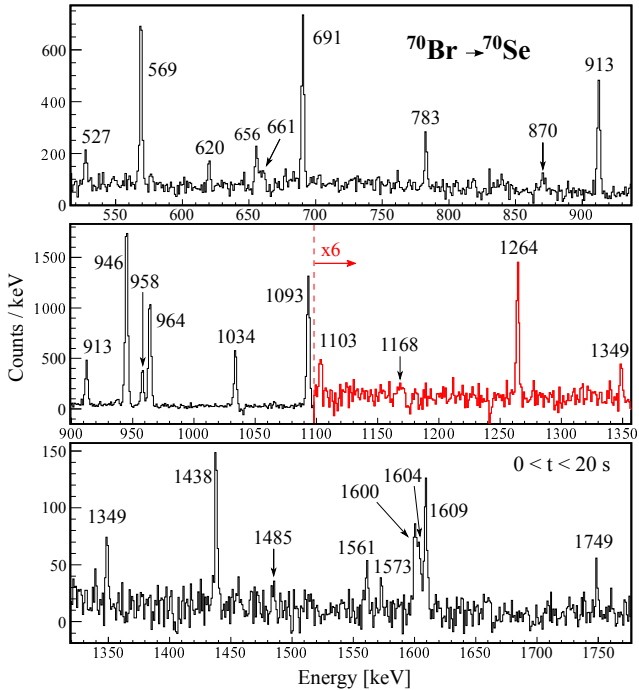}  
\caption{(color on-line) Singles $\gamma$-ray energy spectrum in coincidence with $\beta$ particles detected within 20 s after the implantation of $^{70}$Br ions. Random background has been subtracted using the backward-time correlations described in the text. The $\gamma$ rays attributed to $^{70}$Se are labeled with their energies. In the central panel, the part of the spectrum on the right of the vertical dashed line (shown in red on-line) is scaled by a factor 6 to facilitate the observation of the $\gamma$ rays.}
\label{fig2}
\end{figure}

Examples of $\beta$-decay time spectra for the $9^+$ and $0^+$ isomers are shown in Fig. \ref{fig1}. These were fitted to the Bateman equations \cite{Bat10} to obtain the half-lives. In the fitting procedure we used a $\chi^2$ minimization algorithm for the evaluation of parameters and goodness-of-fit. Because the standard $\chi^2$ test  works properly for histogram data only if both Gaussian and Poisson statistics are applicable, the number of counts in the histogram bins was kept larger than $\sim10$. Thus, we did not subtract the random ion-$\beta$ correlations from the decay curves, but added a constant background function to the fit. The  background was evaluated in separate time-correlated spectra including the so-called backward-time correlations \cite{Kur08,Mor14PRL}, which were built between implantations and preceding $\beta$ particles using the conditions (1) to (3) and (a) to (d) as for the normal time correlations. 

In Fig.~\ref{fig1}, the minimized fitting functions are indicated as continuous lines. It should be noted that at least the first 2 ms were excluded from the fits in order to avoid decay-dependent dead-time contributions due to the electronic processing of implantation events \cite{Goi16}. For the sake of clarity, the contributions from the $0^+$ and $9^+$ activities are indicated as dashed and dotted lines, respectively, while the constant background is shown as a dot-dashed line. The goodness of the fits in describing the data is confirmed by the reduced $\chi^2$ values and the bin-by-bin residual plots, which are also shown in the figure. 
 
 In order to obtain $\beta$-delayed $\gamma$-ray energy spectra, the maximum time elapsed between implantations and $\beta$ particles was fixed to five half-life periods. In addition, the time difference between $\beta$ particles and $\gamma$ rays (henceforth called $\beta(\gamma)$ correlation) was set to 800 ns to include low-energy $\gamma$ rays affected by the electronic time walk, though it could be extended up to 100 $\mu s$ in the presence of isomeric transitions. 
Background contributions from other nuclear species present in WAS3ABi were evaluated in separate $\beta$-delayed $\gamma$ spectra including the backward-time correlations.

\section{$\beta$ decay of the T=0 (J$^{\pi}$=9$^+$) isomer}
\label{sec:long}
\subsection{Half-life}

The half-life of the $T=0~(J^{\pi}=9^+)$ isomer was extracted from the time behavior of the transitions at 945, 964, 1033, and 1093 keV, which were unambiguously assigned to $^{70}$Se in Refs. \cite{Pie00,Roe02,Rai02}. For each peak, spurious correlations with Compton $\gamma$ rays were evaluated by sorting ion-$\beta(\gamma)$ time distributions with well-defined close-lying Compton background. These spectra were normalized to the area under the peak defined by a linear-polynomial background function, and were subtracted from the ion-$\beta(\gamma)$ time-correlated spectrum for each transition. The resulting summed $\gamma$-ray time-decay histogram, divided in time bins of 20 ms, is shown in Fig. \ref{fig1}(a) for an interval of 20 s. The fitting function includes the exponential $\beta$ decay of the $T=0~(J^{\pi}=9^+)$  state and a fixed background extracted from a constant linear fit to the backward-time distribution of ion-$\beta$ correlations. The final half-life is obtained from the average of the half-lives resulting from varying the bin width, the fitting range, and the starting fitting time of the time-correlated spectra. The overall half-life deduced is $t_{1/2}=2157^{+53}_{-49}$ ms. The error results from the quadratic sum of the statistical and systematic uncertainties, which amount to $\pm 30$ ms and $^{+44}_{-38}$ ms, respectively. Note that the systematic error is obtained from the quadratic sum of the uncertainties associated to the fixed background ($^{+24}_{-11}$ ms), the bin width ($\pm 18$ ms), the fitting range ($\pm 8$ ms), and the starting fitting time ($\pm 31$ ms).

\begin{figure}
\centering
\vspace*{0.2cm}
\hspace*{-0.5cm}
\includegraphics[width=8.0 cm]{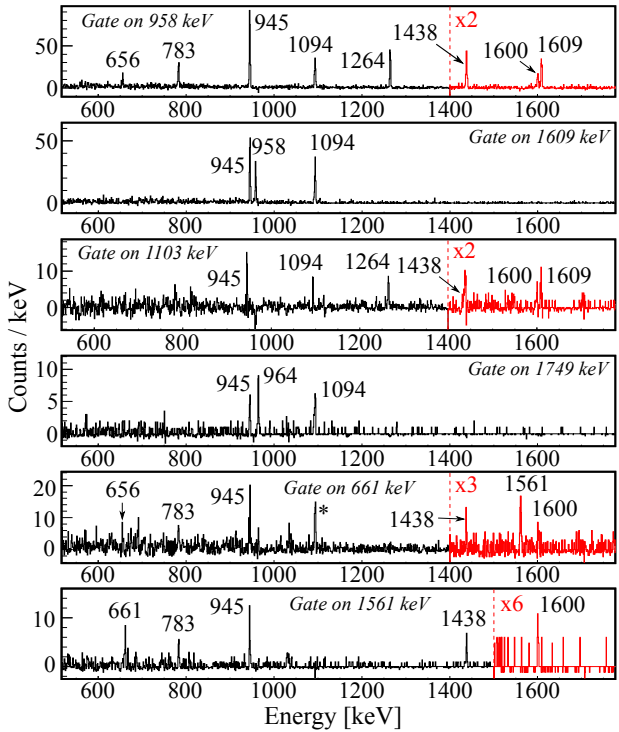}  
\caption{(color on-line) $\beta$-delayed $\gamma$-$\gamma$ coincidence spectra gated on the 958-, 1609-, 1103-, 1749-, 661-, and 1561-keV $\gamma$ rays attributed to the $\beta$ decay of the $T=0~(J^{\pi}=9^+)$ isomer of $^{70}$Br. The parts of the spectra on the right of the vertical dashed lines (shown in red on-line) are scaled by the factors indicated in each line to enhance the low-intensity $\gamma$ rays. The transition marked by an asterisk does not show a mutual coincidence relationship. }
\label{fig3}
\end{figure}

\subsection{$\beta$-delayed $\gamma$ spectroscopy}

\begin{figure*}
\centering
\includegraphics[width=16cm]{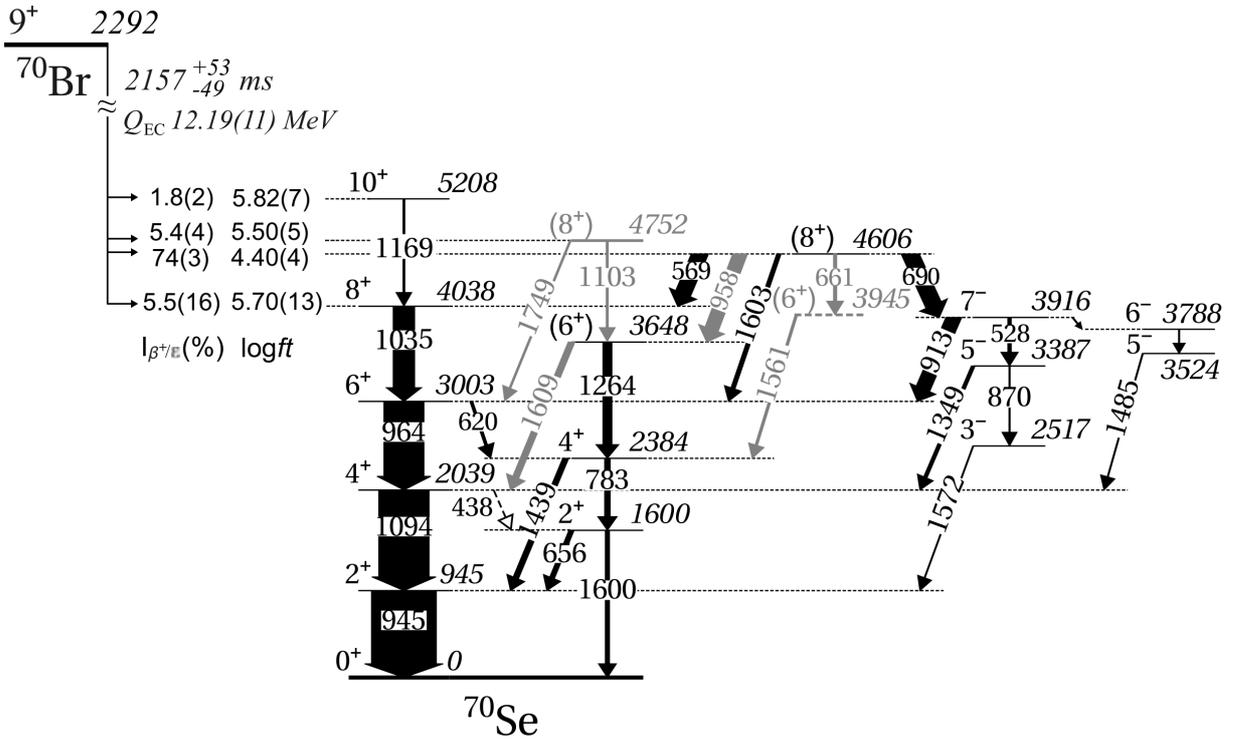}  
\caption{Partial level scheme of $^{70}$Se attributed to the $\beta$ decay of the $T=0~(J^{\pi}=9^+)$ isomer of $^{70}$Br. Spins and parities $J^{\pi}$ of the observed states are indicated on the left side of the levels. Widths of the arrows are proportional to absolute intensities of $\gamma$ rays. New information is indicated in gray. See text for details.}
\label{fig4}
\end{figure*}

The background-subtracted $\beta$-delayed $\gamma$-ray energy spectrum including $\beta(\gamma)$ coincidences within 20~s after implantation of $^{70}$Br ions is shown in Fig. \ref{fig2}. A total of 28 $\gamma$ rays are attributed to the $\beta$ decay of the $T=0~(J^{\pi}=9^+)$ isomer, of which the 661-, 1103-, 1561-, and 1749-keV transitions are reported here for the first time and the rest were reported in previous works \cite{Wad80,Ahm81,Myl89,Pie00,Roe02,Sch02,Rai02}.  The $\beta$($\gamma\gamma$) coincidence spectra gated on the new transitions and on the 958- and 1609-keV $\gamma$ rays are shown in Fig. \ref{fig3}. They are sorted for a maximum $\gamma$-$\gamma$ time difference of 300 ns.  
From these coincidence relations and $\gamma$-ray intensity-balance analysis, the $\beta$-decay scheme for $^{70}$Se is built as shown in Fig. \ref{fig4}. In the figure, arrow widths are proportional to transition intensities, spins and parities are in parentheses if not firmly assigned, and new (or modified) information is indicated in gray. Apparent $\beta$ feedings are extracted from $\gamma$-ray intensity balances. It is assumed that the direct ground-state $\beta$ feeding is null given the large spin difference ($9^+\rightarrow0^+$) between the initial and final states. Hence, the intensities of the $\beta$-delayed $\gamma$ rays are normalized to the summed intensity of the $2^+_{1,2}\rightarrow 0^+$ $\gamma$ transitions. The log$ft$ values are calculated for a $Q_{EC}$ energy of $12.19(11)$ MeV, which was measured using a total absorption spectrometer \cite{Kar04}. Note that any missing $\gamma$ intensity from states located above the observed ones has not been considered, thus $\beta$ feedings and log$ft$ values  should be taken as upper and lower limits, respectively \cite{Har77}.

In general, we find a good agreement with previous $\gamma$-spectroscopy works \cite{Wad80,Ahm81,Myl89,Pie00,Roe02,Sch02,Rai02}, with the exception of the location of the 1609-keV transition which was previously suggested to feed the $2^+_1$ state from a $(4^+)$ candidate at 2553 keV \cite{Myl89}. Looking at our $\beta$-delayed $\gamma$-gated coincidence data in Fig. \ref{fig3}, we attribute this transition unambiguously to the $(6^+_2)\rightarrow4^+_1$ decay. We also confirm the placement of the 958-keV transition as connecting the $(8^+_2)$ and $(6^+_2)$ states as proposed in earlier conference proceedings \cite{Roe02,Sch02}.

Apart from the structure already reported in the literature, two new states are tentatively placed at 3945 and 4752 keV. The location of the 4752-keV level is supported by the observation of two de-exciting transitions, one at 1749 keV feeding the $6^+_1$ state and the other at 1103 keV feeding the $(6^+_2)$ level. Based on this $\gamma$-decay pattern and on the observation of direct $\beta$ feeding from the $9^+$ isomer in $^{70}$Br, we propose a spin and parity $J^{\pi}=(8^+)$ for the new level. On the other hand, the 3945-keV state is indicated by a dashed red line in Fig. \ref{fig4} because the ordering of the 661- and 1561-keV coincident $\gamma$ rays cannot be  established unambiguously from the $\gamma$ intensity balance. We propose a $661\rightarrow1561$-keV $\gamma$ cascade connecting the $(8^+_2)$ and $4^+_2$ states based on comparison with shell-model calculations (see Sec. \ref{sec:discussionb} and Fig. \ref{fig7}). Given its $\gamma$ de-excitation pattern, its most likely spin and parity is $(6^+)$. This assignment is supported by the non-observation of direct $\beta$ feeding from the $9^+$ state in $^{70}$Br.

\section{$\beta$ decay of the T=1 (J$^{\pi}$=0$^+$) isomer}
\label{sec:short}
\subsection{Half-life}

As will be discussed later, no $\gamma$ rays were observed in the decay of the $T=1~(J^{\pi}=0^+)$ isomer (see Fig. \ref{fig6}).  Therefore, the half-life of the superallowed Fermi transition was extracted directly from time correlations between $\beta$ particles and implantations of $^{70}$Br. These are sorted in Fig. \ref{fig1}(b) for a time bin of 1 ms and a correlation time interval of 5 s. The activities of the $9^+$ isomer, daughter ($^{70}$Se) and grand-daughter ($^{70}$As) nuclei were included in the fitting function, together with a fixed background extracted from a constant linear fit to the backward-time correlated spectra. The half-life of the $T=0~(J^{\pi}=9^+)$  state was fixed to the value measured in the present work, $t_{1/2}=2157^{+53}_{-49}$ ms, while those of  $^{70}$Se and $^{70}$As were fixed to the literature values of $2466\pm18$ s and $3156\pm18$ s respectively \cite{ENSDF}. The fixed half-life parameters were varied by one standard deviation to evaluate their contribution to the overall uncertainty. Furthermore, we searched for other factors that could influence the evaluation of the half-life, such as the bin width of the time-correlated spectra, the correlation time interval, the starting time of the fitting range, the $\beta$ threshold, and the ion-$\beta$ correlation strip. The only free parameters in the fit were the half-life of the $T=1~(J^{\pi}=0^+)$ ground state, the number of decays in the first time bin, and the production ratio of the $9^+$ isomer.

\begin{table}[b]
\centering 
 \renewcommand{\arraystretch}{1.4}
 \renewcommand{\tabcolsep}{0.6 cm}

\caption{Error contributions to the half-life determination of the $T=1~(J^{\pi}=0^+)$ ground state of $^{70}$Br. The total uncertainty is given in the last row.} 
\label{table1}
\begin{tabular}{l l }
\hline
\hline
Source & Contribution (ms)  \\  
\hline
bin &  0.10\\
starting time of fit & 0.14 \\
fit range & 0.0013 \\
half-life $^{70}$Br isomer &  0.16\\
half-life $^{70}$Se &  0.002\\
half-life $^{70}$As &  0.002\\
background & 0.10 \\
$\beta$ threshold &  0.11\\
Statistical & 0.43\\

\hline
Total &  0.51\\
\hline
\hline

\end{tabular}

\end{table}

Table \ref{table1} lists the sources of errors evaluated in this work together with their contribution to the total uncertainty in the half-life of the $T=1~(J^{\pi}=0^+)$ ground state. The total error amounts to 0.51 ms after adding in quadrature the listed uncertainties, as indicated in the last row of the table. The overall half-life measured in this work is $t_{1/2}=78.42\pm0.51$~ms, reaching a precision better than $7$ parts in a thousand. The new result is slightly more precise and in excellent agreement with the best value reported thus far in the literature, $t_{1/2}=78.54\pm0.59$ ms \cite{Bur88}. Looking at the second column of table  \ref{table1}, we can see that the main error contribution in this measurement comes from the statistical uncertainty associated to the fit. This is due to the fact that the production ratio of the $9^+$ isomeric state is also a free parameter of the fit. The resulting value is $R^{m}=41.6(8)\%$. 

\begin{figure}
\centering
\vspace*{0.1cm}
\includegraphics[width=8.3 cm]{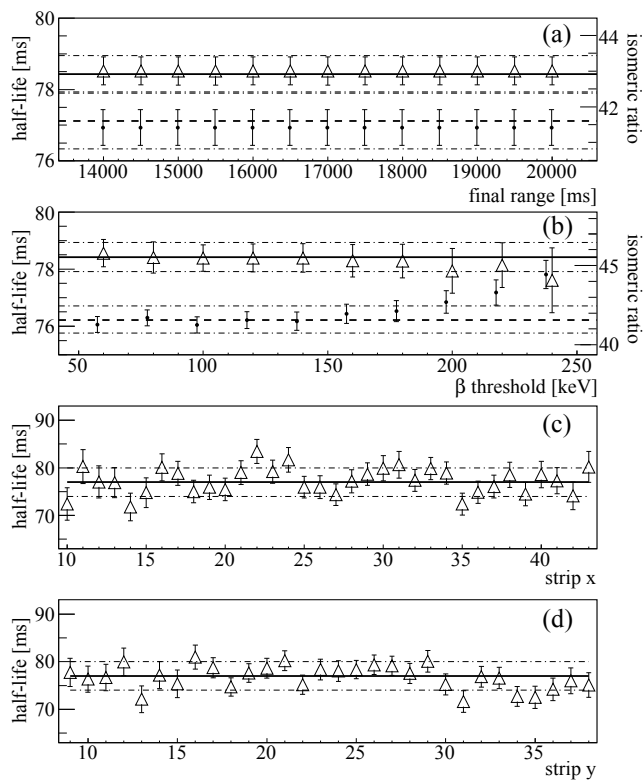}  
\caption{Measured half-lives (empty triangles) for the $T=1$ $(J^{\pi}=0^+)$ ground state as a function of (a) the fitting range, (b) the $\beta$ threshold, (c) the X strips, and (d) the Y strips. The half-lives shown in panels (a), (c), and (d) are obtained without imposing any specific condition on the $\beta$-threshold. Error bars indicate fit uncertainties for each data point. The overall deduced half-life is shown by a thick continuous line, whilst the total error is indicated as a dotted-dashed line. Panels (a) and (b) include in dots the isomeric ratio of the $T=0~(J^{\pi}=9^+)$ state obtained from each lifetime fit. The corresponding  scale is shown on the right axis, and the mean value as a thick dashed line. The dots in panel (b) are slightly  shifted to the left to facilitate their view.}
\label{fig5}
\end{figure}

The half-life of the $T=1~(J^{\pi}=0^+)$ ground state can also be evaluated with an independent procedure which is based on fitting separately the decay-time spectra for each X and Y strip. Accordingly, we can consider $N_x$ or $N_y$ uncoupled measurements of the half-life and calculate the weighted Gaussian mean of the distribution \cite{Har05}. The only limitation comes from the statistical significance of the measurement for each individual strip, which requires, as mentioned above, a number of counts in the histogram bins greater than 10. As a consequence, the strips that do not fulfill this condition are excluded from the analysis (see panels (c) and (d) of Fig. \ref{fig5}). The resulting half-life is $t_{1/2}=77.2\pm 3.0$ ms ($\chi^2_{\nu}=1.107$) in the case of the X-strip analysis and $t_{1/2}=76.8\pm 2.9$ ms ($\chi^2_{\nu}=1.199$) for the Y-strip study. In spite of the rather large uncertainties, the extracted half-lives are in agreement with the result from the analysis of the summed spectra.

In Fig. \ref{fig5}, the half-life of the $T=1~(J^{\pi}=0^+)$ ground state is shown as a function of (a) the fitting range, (b) the $\beta$ threshold, (c) the X strips, and (d) the Y strips. The half-lives shown in panels (a), (c), and (d) are obtained without imposing any specific condition on the $\beta$-threshold. Error bars indicate fit uncertainties for each data point. The average half-life deduced is shown as a thick continuous line and the total error as a dotted-dashed line. Panels (a) and (b) also show, in dots, the isomeric ratio of the $T=0~(J^{\pi}=9^+)$ state for each lifetime fit and, in thick dashed line,  the overall deduced value. For the sake of clarity, the corresponding scale is indicated on the right axis. In all cases the half-lives and isomeric ratios fluctuate statistically around the mean values, except for  $\beta$ thresholds above 200 keV for which the results are out of the error bars. This is because the statistics drop significantly for these data points (note that the associated errors are bigger and, in the case of the half-life, consistent with the overall mean value). The flat behavior of the half-life in panel \ref{fig5}(a) indicates that no contaminant activities apart from those already included in the fit are present in the time-correlated spectra.  


\subsection{Branching ratio through the $2^+_1$ level }

The Gamow-Teller (GT) branching ratios in the $\beta$ decay of selected $62\leq A\leq 74$ nuclei were calculated by J. C. Hardy and I. S. Towner using the shell model \cite{Har02}. Because the $Q_{EC}$ values of these heavy nuclei are large, the total GT branching fraction is the sum of many individual weak GT transitions. In the case of  $^{70}$Br, a total of 325 $1^+$ states are expected to be fed through GT $\beta$ transitions, resulting in a GT branching ratio of $1.59\%$ \cite{Har02}. In the calculation, $63\%$ of the GT intensity is expected to de-excite through the $2^+_1$ level, resulting in $R^{GT}(2^+_1)=1.59\times0.63\approx1\%$. 

Experimentally, non-analog branching ratios (including both GT and non-analog Fermi $\beta$ strength) are usually obtained from the intensities of the observed $\gamma$ rays feeding the ground state. Because in some cases the strong fragmentation of the $\beta$ feeding prevents a direct observation of the de-exciting transitions \cite{Har77}, an estimate can be obtained from the measured $\gamma$ imbalance of  the $2^+_1$ level, which acts as a collector of an important part of the non-analog intensity \cite{Pie03}.

\begin{figure}
\centering
\vspace*{0.2cm}
\hspace*{-0.5cm}
\includegraphics[width=8.5 cm]{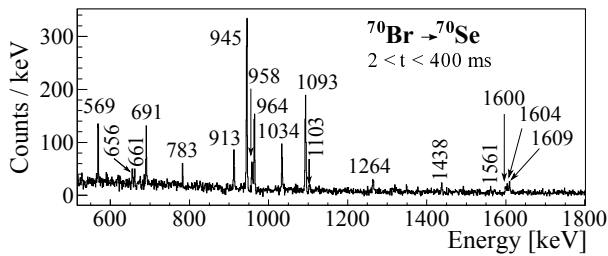}  
\caption{$\beta$-delayed $\gamma$ energy spectrum following the implantation of $^{70}$Br ions. Though the time interval 2-400 ms is selected to enhance the decay of the $T=1~(J^{\pi}=0^+)$ ground state, the only $\gamma$ rays observed are attributed to the decay of the $9^+$ isomer.}
\label{fig6}
\end{figure}

Figure \ref{fig6} shows the $\beta$-delayed $\gamma$-ray spectrum sorted in a time interval $2-400$ ms after the detection of $^{70}$Br implantations to enhance $\gamma$ rays emitted following the decay of the $T=1~(J^{\pi}=0^+)$ state. All the observed $\gamma$ transitions were identified before as resulting from the decay of the high-spin isomer. Yet we can estimate the fraction of the branching ratio from the $T=1~(J^{\pi}=0^+)$ ground state that decays to the ``collector'' $2^+_1$ level from the measured intensity of the $2^+_1\rightarrow 0^+_1$ 945-keV $\gamma$ transition in two variable time windows. The first, $\Delta t_1$, contains the decay of both the high-spin isomer and the ground state, while the second, $\Delta t_2$, includes only the decay of the high-spin isomer. In order to let the $T=1~ (J^{\pi}=0^+)$ activity exhaust, we constrain the starting time of the second time interval to, at least, 15 half-life periods. 

The number of counts in the 945-keV peak in each of the two time intervals can be expressed as

\begin{equation}
 \centering
 N_{\gamma}(\Delta t_2) =N^{\beta}_{0}\cdot F(\Delta t_2,9^+)\cdot R^{m}\cdot\varepsilon_{\gamma}\cdot I_{\gamma}^{T=0}(2^+_1)
 \label{eq:eq2}
\end{equation}

\begin{equation}
 \centering
 \begin{split} 
 N_{\gamma}(\Delta t_1) = N^{\beta}_{0}\cdot F(\Delta t_1,9^+)\cdot R^{m}\cdot\varepsilon_{\gamma}\cdot I_{\gamma}^{T=0}(2^+_1) \\
+ ~ N^{\beta}_{0}\cdot F(\Delta t_1,0^+)\cdot (1-R^{m})\cdot\varepsilon_{\gamma}\cdot I_{\gamma}^{T=1}(2^+_1) \\
 \end{split}
  \label{eq:eq3}
\end{equation}

where $N^{\beta}_{0}$ is the total number of decays, $F(\Delta t,J^{\pi})$ is a correction factor that takes into account the finite time of the measurement (it is defined as the ratio between the activity integrated in the time-correlation window and an infinite time window), $R^m$ is the production ratio of the  $T=0~ (J^{\pi}=9^+)$ isomer in the fragmentation reaction, $I_{\gamma}^{T}$ is the absolute intensity of the $2^+_1\rightarrow 0^+_1$ transition following $\beta$ decay of the $T=0$ or 1 isomers, and $\varepsilon_{\gamma}$ is its absolute photopeak efficiency.  Note that the internal conversion coefficient, $\alpha_{tot}$, of the 945-keV transitions is not taken into account in expressions \ref{eq:eq2} and \ref{eq:eq3}. This is a good approximation given the calculated value for a pure $E2$ transition, $\alpha_{tot}=4.81(7)\times10^{-4}$ \cite{Kib08}. Furthermore, no dead-time corrections are included since $\gamma$-ray information was only recorded in coincidence with accepted $\beta$-like signals from WAS3ABi.   

From expressions \ref{eq:eq2} and \ref{eq:eq3} we can estimate the total branching fraction from the $T=1~(J^{\pi}=0^+)$ ground state in $^{70}$Br that decays through the $2^+_1$ level  in $^{70}$Se. The resulting value, $R(2^+_1)= I_{\gamma}^{T=1}(2^+_1)=1.3\pm1.1\%$, can be combined with the GT feeding accumulated by the $2^+_1$ state as calculated in Ref. \cite{Har02},  $R^{GT}(2^+_1)\approx1\%$, to estimate the ratio between the non-analog Fermi and Gamow-Teller feedings decaying through the $2^+_1$ state in $^{70}$Se, $R^{F/GT}=30\%$. This value is in between the ones measured for the neighboring $N=Z$ nuclei $^{62}$Ga and $^{74}$Rb, $R^{F/GT}\sim10\%$ \cite{Fin08} and $R^{F/GT}\sim57\%$ \cite{Dun13}, respectively.  

\section{Discussion}
\label{sec:discussion}
\subsection{Testing CVC with the present results}

\begin{table*}
\centering 
\renewcommand{\arraystretch}{1.4}
\renewcommand{\tabcolsep}{0.34 cm}

\caption{Determination of $\mathcal{F}t$ values for the superallowed decay of $^{70}$Br. First line, using the $Q_{EC}$ from \cite{Dav80,Har15},  the $f$ value quoted in \cite{Har15} and the experimental $t_{1/2}$ and $R$ determined in this work. Second line, using the $Q_{EC}$ value from \cite{Sav09,Wan17} and the $f$ value determined using the LOGFT code available in the NNDC website \cite{NNDC}. Transition-dependent radiative and nuclear corrections are taken from Ref. \cite{Har15}, $\delta'_R = 1.49\%$ and $\delta_C-\delta_{NS}= 1.78\pm0.25\%$.} 
\label{table2}
\begin{tabular}{ c c c c c c c}
\hline
\hline
$Q_{EC}$ (keV) & $t_{1/2}$ (s)  & $f$  &  $R$  (\%)   & $P_{EC}$   (\%) & $ft$  (s) & $\mathcal{F}t$ (s)  \\  
\hline
$9970\pm170$ \cite{Dav80} & $0.07842\pm0.00051$ & $38600\pm3600$ \cite{Har15} & $97.94\pm1.75$ & 0,173 & $3096\pm293$ & $3086\pm293$ \\
$10504\pm15$ \cite{Sav09} & $0.07842\pm0.00051$ & $50979\pm385$ \cite{NNDC} & $97.94\pm1.75$ & 0,133 & $4087\pm83$ & $4078\pm83$ \\

\hline
\hline

\end{tabular}

\end{table*}

The fraction of the branching ratio going through the $2^+_1$ level obtained in the previous section can be combined with the shell-model predictions of Ref. \cite{Har02} to estimate the superallowed $0^+_{gs}\rightarrow0^+_{gs}$ branching fraction of $^{70}$Br. According to the theoretical result, 63\% of the summed GT strength goes through the  $2^+_1$ level. If we assume that the ratio between the $\gamma$ intensities decaying from non-analog Fermi and Gamow-Teller states is the same for the ground state, we can estimate a total non-analog branch $R^{na}=2.06\pm1.75\%$ for this decay. This results in a superallowed $0^+_{gs}\rightarrow0^+_{gs}$ branching ratio $R=97.94\pm1.75\%$.

Both the half-life and the $0^+_{gs}\rightarrow0^+_{gs}$ branching fraction discussed in this work have been used to estimate the $\mathcal{F}t$ value associated with the decay of the $T=1~(J^{\pi}=0^+)$ state of $^{70}$Br. This allows one to check how well it agrees with the average $\overline{\mathcal{F}t}$ value obtained from the best 14 cases in the last compilation of superallowed $0^+\rightarrow 0^+$ transitions \cite{Har15}. In the determination of ${\mathcal{F}t}$ we have followed the procedure outlined in Ref. \cite{Har15}, which implies obtaining the partial half-life $t$ using the following formula:
\begin{equation}
t=\frac{t_{1/2}}{R}(1+P_{EC})
\end{equation}
where $t_{1/2}$ is the half-life of the parent state, $R$ is the branching ratio of the superallowed transition, and $P_{EC}$ is the electron-capture fraction. 

The values of $Q_{EC}$, $t_{1/2}$, $f$, $R$ and $P_{EC}$ employed in the calculation are presented in Table \ref{table2} together with the uncorrected $ft$ and corrected $\mathcal{F}t$ values. In the first line of the table we show the $\mathcal{F}t$ value calculated using the total transition energy obtained in a measurement of the positron end-point energy, $Q_{EC}$=9970(170) keV \cite{Dav80}, which was proposed in the last compilation of superallowed Fermi transitions \cite{Har15}. In this case we have used the $f$ value calculated by Ref. \cite{Har15}. In the second line of the table we have determined the $\mathcal{F}t$ value for a total transition energy $Q_{EC}$=10504(15) keV, which was adopted in the last atomic mass evaluation \cite{Wan17} from a Penning-trap mass measurement \cite{Sav09}. In this case the $f$ value has been determined using the LOGFT code available in the National Nuclear Data Center (NNDC) website \cite{NNDC}. 

Looking at the table, the $\mathcal{F}t$ value determined in the first row, $\mathcal{F}t=3086\pm293$ s, is in good agreement with the last average value,  $\overline{\mathcal{F}t}=3072.27\pm0.72$ s \cite{Har15}. On the other hand, the $\mathcal{F}t$ value obtained in the second row, $\mathcal{F}t=4074\pm83$ s, differs dramatically from the mean $\overline{\mathcal{F}t}$, pointing to a possible incorrect determination of the $Q_{EC}$ value from the masses measured in the Penning-trap experiment \cite{Sav09}. The $Q_{EC}$ value of this decay is specifically discussed in Ref. \cite{Har15} because it does not follow the trend of the rest of the $Q_{EC}$ values for $T_{Z}$=0 superallowed transitions. In addition, it is in clear conflict with the CVC hypothesis, as shown by the resulting ${\mathcal{F}t}$ value listed in Table \ref{table2}. In order to achieve a better sensitivity for the CVC test, both the $Q_{EC}$ value and the branching fraction of the superallowed transition in $^{70}$Br should be measured with improved uncertainties. 

\subsection{Comparison with Shell-Model Calculations}
\label{sec:discussionb}

\begin{figure*}
\centering
\includegraphics[width=15 cm]{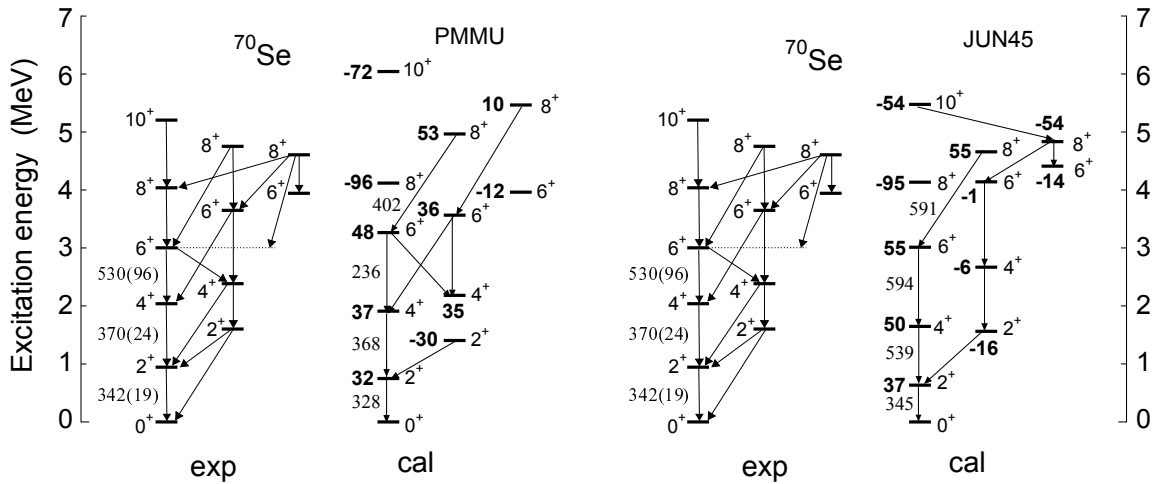}  
\caption{Experimental (exp) and theoretical (cal) low-lying level schemes of $^{70}$Se. Shell-model calculations using the PMMU \cite{Kan14} (left) and JUN45  \cite{Hon09} (right) interactions are shown. Spectroscopic quadrupole moments are indicated in units of e$\cdot$fm$^2$ on the left side of the calculated states, in boldface. Arrows in the experimental level scheme indicate observed electromagnetic transitions, while in the theoretical ones they stand for states connected by large B(E2$\downarrow$) values. Experimental \cite{Lju08} and theoretical B(E2$\downarrow$) values are shown next to the corresponding transitions.}
\label{fig7}
\end{figure*}

The structure of $^{70}$Se has been interpreted using large-scale shell-model calculations performed in the $p$+$f_{5/2}$+$g_{9/2}$ model space. Both the effective interactions PMMU \cite{Kan15} and JUN45 \cite{Hon09} have been employed. The former is based on the unified realistic shell-model Hamiltonian PMMU \cite{Kan14}, which includes the pairing-plus-multipole Hamiltonian and a monopole interaction extracted from empirical fits starting from the monopole-based universal force. The latter is derived from the realistic Bonn-C nucleon-nucleon potential and a Linear Combination fit \cite{Chu76} to 400 experimental energies of 69 nuclei with $A=63-96$. Both interactions have successfully been used to describe the nuclear properties of $N\approx Z$ nuclei in the $A=64-80$ mass region \cite{Kan14, Kan15,Hon09}. In the present calculations, the effective charges for proton and neutron have been taken as $e_p=1.5$e and $e_n=0.5$e, respectively, which provide a good agreement with the observed B(E2) values in $A\sim70$ nuclei \cite{Lut12,Nic14}. Furthermore, the PMMU Hamiltonian has been modified so as to fit the higher spin states of $^{70}$Se. 

In Fig. \ref{fig7} the experimental low-lying structure of $^{70}$Se is compared to the level schemes calculated using the PMMU (left) and JUN45 (right) interactions. Arrows connecting experimental levels stand for electromagnetic transitions observed in the present work, while arrows connecting calculated levels indicate large theoretical B(E2$\downarrow$) values. Spectroscopic quadrupole moments are also shown on the left of the theoretical states. They are expressed in units of e$\cdot$fm$^2$. 

In general, the excitation energies of the observed levels are well reproduced by both calculations, with the PMMU having a better accuracy for spins up to $6^+$ and the JUN45 for the $8^+$ and $10^+$ states. Similarly, the calculated spectroscopic quadrupole moments are in good agreement, except for the second $4^+$ and $6^+$ levels and the third $8^+$ state for which JUN45 predicts negative values and PMMU positive ones. Note that the experimental $(8^+_3)$ level has been placed in the band built on the  $2^+_2$ state. This is due to the non-observation of a $(8^+_3)\rightarrow (6^+_3)$ $\gamma$ transition connecting the two newly observed states. Instead, the $(6^+_3)$ level is only fed by the $(8^+_2)$ state at 4606 keV. This results in an inversion of the $(8^+)$ states of bands 2 and 3. 

Both calculations predict a prolate-deformed shape for the yrast $8^+$ level, as indicated by the large negative spectroscopic quadrupole moments $Q_s$ shown in Fig. \ref{fig7}. On the contrary, the yrast states up to $J^{\pi}=6^+$ have positive $Q_s$ values and can be interpreted as being dominated by oblate-deformed configurations. These calculations are consistent with the experimental evidence found thus far in the region: previous Coulomb excitation \cite{Hur07} and recoil-distance Doppler shift \cite{Lju08} measurements revealed a positive sign for the quadrupole moment of the first $2^+$ state in $^{70}$Se, thus confirming its oblate shape. As well, spectroscopic works on the neighboring $^{69,71}$Se provided good experimental evidence for oblate deformation in the low-lying levels of the $g_{9/2}$ band \cite{Wio88}. On the other hand, the $9^+$ state of $^{70}$Br is predicted to have a large negative $Q_s$ value ($Q_s=-120$ e$\cdot$fm$^2$ for PMMU and $Q_s=-117$ e$\cdot$fm$^2$ for JUN45), which indicates strong prolate deformation as for the  yrast $8^+$ states in $^{70}$Se. The calculated Gamow-Teller transition strength to these levels results in B$(GT)=0.156$ and log$ft=4.62$ for PMMU and B$(GT)=0.131$ and log$ft=4.69$ for JUN45. These results compare well with log$ft=4.40(4)$ measured in the present work for the $(8^+_2)$ level, indicating that they can be interpreted as their theoretical counterparts.

Experimentally, the $(8^+_2)$ state is connected to the $8^+_1$ and $(6^+_2)$ levels by intense transitions of 569- and 958-keV energy, respectively. This  indicates a strong overlap of the wave functions of the three states and, hence, of configuration mixing. Instead, the 1603-keV $\gamma$ ray decaying to the yrast $6^+_1$ state is suppressed by a factor $\sim$3 with respect to the $(8^+_2)\rightarrow (6^+_2)$ transition at 958-keV even if it is energetically favored by a factor $\sim$13. This suggests a change in the wave functions of the $(8^+_2)$ and $(6^+_2)$ states with respect to the $6^+_1$ level that is theoretically supported by the JUN45 calculation, which predicts differing prolate and oblate characters for them. This is not the case for the PMMU calculation, for which the $6^+_2$ state is predicted to have a positive $Q_s$ value and then a mainly oblate character. Our results can also be compared with state-of-the-art  nuclear models such as the complex Excited VAMPIR variational approach  \cite{Pet03,Pet15} and the adiabatic self-consistent collective coordinate (ASCC) method \cite{Hin09}, which predict a strong mixing between oblate and prolate configurations in the yrast band up to spin $J\sim4$ and a clear dominance of prolate configurations at higher spins. 

As a final remark, one should note that the experimental and theoretical results reported thus far show a rather complex shape-coexistence picture at low excitation energies in $^{70}$Se. The main missing piece in this puzzle is the first excited $0^+$ state predicted at a low excitation energy by the shell model \cite{Kan15}. Its observation would provide strong evidence for shape coexistence and clarify largely the present picture. 

\section{Summary and Conclusions}
\label{sec:conc}

The $\beta$ decays of the $T=1~(J^{\pi}=0^+)$ and $T=0~(J^{\pi}=9^+)$ isomers in $^{70}$Br have been simultaneously investigated at the RIBF facility of the RIKEN Nishina Center using the BigRIPS and EURICA setups. Improved values are obtained for the half-lives of both states, $t_{1/2}(9^+)=2157^{+53}_{-49}$ ms and  $t_{1/2}(0^+)=78.42\pm51$ ms. In the case of the $T=1~(J^{\pi}=0^+)$ ground state, an estimate of the total branching fraction decaying through the $2^+_1$ state, $R(2^+_1)=1.3\pm1.1\%$, is provided for the first time. These results have been combined with the shell-model predictions of2 Ref. \cite{Har02} to test the CVC hypothesis in the heavy self-conjugate nucleus $^{70}$Br. We have confirmed that the corrected $\mathcal{F}t$ value for the Penning-trap mass measurement accepted in the last atomic mass evaluation \cite{Sav09,Wan17} does not satisfy CVC, while that for an old $\beta$ end-point measurement  \cite{Dav80,Har15} does. At present, this  $Q_{EC}$ value is the major contributor to the large uncertainty in the $\mathcal{F}t$ estimate for $^{70}$Br. In order to improve the sensitivity of the CVC test for this nucleus, a new mass measurement is needed. In addition, improved error budgets for the half-life and the superallowed branching fraction are desirable. 

In the decay of the $T=0~(J^{\pi}=9^+)$ isomer, two new excited states at  3945 and 4752 keV with $J^{\pi}=(6^+)$ and $(8^+)$ have been proposed for the first time. Their nature has been discussed in terms of large-scale shell-model calculations including the PMMU \cite{Kan15} and JUN45 \cite{Hon09} interactions for the $p_{3/2}$, $p_{1/2}$, $f_{5/2}$, and $g_{9/2}$ configuration space. For most of the levels the excitation energies are well reproduced by the calculations. Theoretical B$(GT)$ values have also been calculated from the prolate-deformed $9^+$ state in $^{70}$Br to the yrast $8^+_1$ level in $^{70}$Se, resulting in log$ft=4.62$ for PMMU and log$ft=4.69$ for JUN45. These results are in agreement with the experimental log$ft=4.40(4)$ to the $(8^+_2)$ state at 4606 keV, suggesting that the latter is the corresponding prolate experimental state. Based on this and on the internal de-excitation pattern of the $(8^+_2)$ level, we propose a slightly modified shape-evolution picture in $^{70}$Se in which the oblate-to-prolate transition in the yrast band might take place more slowly than previously expected.

\section*{Acknowledgements}

We want to express our gratitude to the RIKEN accelerator staff for providing a stable and high intensity $^{78}$Kr beam during the experiment, the EUROBALL Owners Committee for the loan of germanium detectors, and the PRESPEC Collaboration for the readout electronics of the cluster detectors. Part of the WAS3ABi was supported by the Rare Isotope Science Project, which is funded by the Ministry of Science, ICT and Future Planning (MSIP) and National Research Foundation (NRF) of Korea. Support from the Spanish Ministerio de Econom\'ia, Industria y Competitividad under Grants No. IJCI-2014-19172, No. FPA2011-24553  and  No.  FPA2014-52823-C2-1-P, Centro de Excelencia Severo Ochoa del  IFIC  SEV-2014-0398, and Junta para la Ampliaci\'on de Estudios Programme (CSIC JAE-Doc contract) cofinanced by FSE is acknowledged. This work was also financed by Programmi di Ricerca Scientifica di Rilevante Interesse Nazionale (PRIN) number 2001024324\_01302, by JSPS KAKENHI Grant No. 25247045, by the Chilean FONDECYT projects INICIACION No.11130049 and REGULAR No. 1171467, by the Istanbul University Scientific Project Unit under Projects No. BYP-53195, by the Max-Planck Partner group, by the NSF Grant No. PHY-1404442, by the UK Science and Technology Facilities Council (STFC) Grant No. ST/F012012/1, by the Japanese Society for the Promotion of Science under Grant No. 26 04808, by MEXT Japan under Grant No. 15K05104, and by the French-Japanese ``Laboratoire international associ\'e'' FJ-NSP. D.Zs. acknowledges the financial support through the GINOP-2.3.3-15-2016-00034 project.

\bibliographystyle{apsrev}
\bibliography{bibliografia_ale}

\end{document}